\documentclass[aps,superscriptaddress,twocolumn,showpacs,preprintnumbers,amsmath,amssymb,wasysym]{revtex4}
\usepackage[cp1251]{inputenc}
\usepackage{amssymb,amsmath,wasysym}
\usepackage[mathscr]{eucal}
\usepackage{graphicx}% Include figure files
\usepackage{longtable}
\usepackage[dvips]{hyperref} 
\usepackage[cp1251]{inputenc}
\usepackage[english]{babel}
\usepackage{amssymb,amsmath}
\usepackage{amstext} 
\usepackage[dvips]{hyperref}
\usepackage[mathscr]{eucal}
\usepackage{longtable}
\usepackage{graphicx}% Include figure files
\begin{document}
\title{Quantum  Nature  of Relaxation of Paramagnetic and Optical Systems by Strong Dipole-Photon and Dipole-Phonon Coupling}
\author{D.Yearchuck (a),  A.Dovlatova (b), V.Stelmakh  (c)\\
\textit{(a) - Minsk State Higher Aviation College, Uborevich Str., 77, Minsk, 220096, RB, E-mail: yearchuck@gmail.com \\(b) - M.V.Lomonosov Moscow State University, Moscow, 119899 \\(c)- Belarusian State University, Nezavisimosti Ave., 4, Minsk, 220030, RB}}
\date{\today}% It is always \today, today,
\begin{abstract}  Matrix-operator difference-differential equations for  dynamics of spectroscopic transitions in 1D multiqubit  exchange coupled  (para)magnetic and optical systems by strong dipole-photon and dipole-phonon coupling are derived within the framework of quantum electrodynamics and quantum phonon field theory.  It has been established, that in the model considered the  relaxation processes  are of pure quantum character, which is determined by  the formation of the coherent system of the resonance phonons and by the appearence along with absorption process of EM-field energy the coherent emission process, acompanying by phonon Rabi quantum oscillation,  which can be time-shared. For the case of radiospectroscopy it corresponds to the possibility of the simultaneous observation  along with (para)magntic spin resonance the acoustic spin resonance. Theoretical conclusions are confirmed experimentally in  radiospectroscopy. It has been found in particular, that the lifetime of coherent state of collective subsystem of resonance phonons in disordered carbon samples - anthracites of medium-scale metamorphism - is very long and even by room temperature it is evaluated in $\sim 10$ ms. The phenomenon of  the formation of the coherent system of the resonance phonons can be used in a number of practical applications, in particular by elaboration of logic quantum systems including quantum computers and quantum communication systems.     
\end{abstract}
\pacs{42.50.Ct, 61.46.Fg, 73.22.–f, 78.67.Ch, 77.90.+k, 76.50.+g}% PACS, the Physics and Astronomy
\maketitle 
\section{Introduction and Background}
Quantum electrodynamics (QED) and quantum field theory at all take on more and more significance for its practical application and thy, in fact, become to be working instrument in spectroscopy studies and industrial spectroscopy control. 
Quantum dynamics of two level systems (qubits),  coupled to a single mode of
an electromagnetic cavity are of considerable interest in 
connection with  quickly developing new quantum physics branches like to cavity
quantum electrodynamics \cite{Berman} or  quantum computing \cite{Nielsen}. Spin-based solid state quantum
bits  are known to have long coherence times,
while also offering the promise of scalability, and are natural
building blocks for quantum computation. Phosphorus donor nuclei in silicon have been known
since the 1950s to have some of the best spin coherence
properties in solids. The spin coherence time $T_2$ measured by   Hahn spin echo method for
donor electron spins in bulk Si:P has been reported to be equal
$\sim 60 ms$ \cite{Tyryshkin}. This is the longest coherence
time measured in electron spin qubits, and greatly
exceeds the value reported, for instance, in GaAs quantum dots, which, measured also by   Hahn spin echo method, is equal to $\sim 100 \mu$s \cite{Koppens}.
 However, fabrication of ordered
and gated donor arrays and coherent control over
donor electrons has turned out to be extremely difficult. There is now much interest in fabricating
spin-based devices in diamond, with potential applications in quantum communication, quantum computation, and magnetometry.
Nitrogen-vacancy (NV) centers located deep within
a diamond lattice appear promising  solid-state spin
qubits since they combine optical initialization and readout capabilities owing to long electron spin coherence life
times approaching $1 ms$ at room temperature \cite{Gaebel}, and the
ability to control coupling to individual nuclear spins.
In all of aforesaid
applications it is necessary or at least advantageous to
couple NV centers to optical structures like to waveguides and microresonators, to enable
communication between distant qubits or to allow efficient extraction of emitted photons. Therefore, a reliable
method is needed to create NV centers with good spectral
properties in close proximity $(\apprle 100 nm)$ to a diamond
surface. In addition, the charge state of the NV center
must be controlled. Given tasks are under study at present.
We will show in present work another way to obtain long-lived coherent states with similar field of practical application. The prediction will be based on QED-theory of spectroscopic transitions.
The simplest models which capture the
salient features of the relevant physics in given  field are
the Jaynes-Cummings model (JCM) \cite{Jaynes_Cummings} for the one qubit
case and its generalization for multiqubit systems by
Tavis and Cummings \cite{Tavis}. Tavis-Cummings model was generalized in \cite{Slepyan_Yerchak}, by taking into account the 1D-coupling between qubits. Recently QED-model for  one chain coupled qubit system was generalized 
  for quasionedimensional axially symmetric  multichain coupled qubit system  \cite{Dovlatova A_Yearchuck D}. 
  It is substantial, that in the model, proposed in \cite{Dovlatova A_Yearchuck D} the interaction of quantized EM-field with multichain qubit system  is considered by taking into account both the intrachain and interchain qubit coupling without restriction on their values.   It follows from theoretical results in \cite{Slepyan_Yerchak}, \cite{Dovlatova A_Yearchuck D} and from their experimental confirmation in 
 \cite{Yearchuck D_Dovlatova A}, that by strong interaction of EM-field with matter the correct description of spectroscopic transitions including stationary spectroscopy is achieved the only in the frame of QED consideration. It  concerns both optical and radio spectroscopies, that means, that QED consideration has to be also undertaken by
 electron spin resonance (ESR) studies in the case of strong interaction of EM-field with spin systems. It is reasonable to suggest, that  analogous conclusion can be
drawn for the case of strong interaction of phonon field with spin system or electron system. In other words it seems to be reasonable the idea, that relaxation of paramagnetic (or optical) centers in the case of strong spin-phonon (electron-phonon) interaction can be described correctly the only in the frames of quantum field theory.

The aim of the work presented is to derive the system of equations for  dynamics of spectroscopic transitions in 1D multiqubit  exchange coupled  (para)magnetic and optical systems by strong dipole-photon and dipole-phonon coupling  within the framework of quantum electrodynamics and quantum phonon field theory (QPTT) and to  show both theoretically and experimentally, that new quantum plysics phenomenon - the formation of longlived coherent state of resonance phonons leading  to appearance uf quantum acoustic Rabi oscillations takes place and to propose the area of its application. 

 \section{Results and Discussion}
\subsection{Quantum Equations for Spectroscopic Transitions by Strong Electron-Photon and Electron-Phonon Coupling}
Recently in the work \cite{Yearchuck_Yerchak_Dovlatova} the system of  difference-differencial equations for dynamics of spectroscopic transitions for both radio- and optical spectroscopy for the model, representing itself the 1D-chain of N two-level  equivalent elements coupled by exchange interaction (or its optical analogue for the optical transitions) betweenn themselves and interacting with quantized EM-field and   quantized phonon field its optical  has recently been  derived. Naturally the equations are true for any 3D system of paramagnetic centers (PC) or optical centers  by the absence of exchange interaction. In given case the model presented  differs from Tavis-Cummings model \cite{Tavis} by inclusion into consideration of quantized phonon system, describing the relaxation processes from quantum fied theory position.  Seven equations for the seven operator variables, describing joint system \{field + matter\} can be presented in matrix form by three matrix equations. They are the following

\begin{equation}
\label{eq1}
\begin{split}
\raisetag{40pt}
\frac{\partial}{\partial t} 
\left[\begin{array}{*{20}c}
{\hat\sigma^-_l}  \\
 \\
{\hat\sigma^+_l}  \\
\\
{\hat\sigma^z_l} 
\end{array} 
\right] = 2\left\|g\right\|\left[\begin{array}{*{20}c}{\hat F^-_l}  \\
 \\
{\hat F^+_l}  \\
\\
{\hat F^z_l} 
\end{array} 
\right] +  ||\hat{R}^{(\lambda)}_{\vec{q}l}||, 
\end{split}
\end{equation}

\begin{equation}
\label{eq2}
\begin{split}
\raisetag{40pt}
&\frac{\partial}{\partial t} 
\left[\begin{array}{*{20}c}
 {\hat{a}_{\vec k^{}}} \\
 \\
 {\hat{a}_{\vec k^{}}^+} \\
\end{array} 
\right] = -i \omega_{\vec k^{}} ||\sigma_P^z|| \left[\begin{array}{*{20}c}
 {\hat{a}_{\vec k^{}}} \\
 \\ 
 {\hat{a}_{\vec k^{}}^+} \\
\end{array} 
\right] \\
\\
& + \frac{i}{\hbar}
\left[\begin{array}{*{20}c}
{-\sum\limits_{l = 1}^N (\hat\sigma_l^{+} + \hat\sigma_l^{-}) v_{l \vec k}^*} \\
\\
{\sum\limits_{l = 1}^N (\hat\sigma_l^{+} + \hat\sigma_l^{-}) v_{l \vec k}} \\
\end{array} \right],
\end{split}
\end{equation}

\begin{equation}
\label{eq3}
\begin{split}
\raisetag{40pt}
\frac{\partial}{\partial t} 
\left[
\begin{array}{*{20}c}
 {\hat{b}_{\vec k^{}}} \\
 \\
 {\hat{b}_{\vec q^{}}^+} \\
\end{array} 
\right] = -i \omega_{\vec q^{}} ||\sigma_P^z|| \left[\begin{array}{*{20}c}
 {\hat{b}_{\vec q^{}}} \\
 \\ 
 {\hat{b}_{\vec q^{}}^+} \\
\end{array} 
\right] 
 + \frac{i}{\hbar}
\left[
\begin{array}{*{20}c}
{-\sum\limits_{l = 1}^N  \hat\sigma_l^{z} \lambda_{\vec q l}} \\
\\
{\sum\limits_{l = 1}^N \hat\sigma_l^{z} \lambda_{\vec q l}} \\
\end{array} \right],
\end{split}
\end{equation}
where
\begin{equation}
\label{eq4}
\begin{split}
\raisetag{40pt}
\left[\begin{array}{*{20}c}
{\hat\sigma^-_l}  \\
 \\
{\hat\sigma^+_l}  \\
\\
{\hat\sigma^z_l} 
\end{array} 
\right] = \hat{\vec{\sigma}}_l =  \hat\sigma^-_l  \vec e_ +  +  \hat\sigma^+_l \vec e_ - +  \hat\sigma^z_l\vec e_z
\end{split}
\end{equation} is  vector-operator of spectroscopic transitions for $l$th chain unit, $l = \overline{2,N-1}$ \cite{Yearchuck_Yerchak_Dovlatova}.
Its components, that is,  the operators 
\begin{equation}
\label{eq6a}
{\hat\sigma_v}^{jm} \equiv {\left|j_v \right\rangle} {\left\langle m_v \right|} 
\end{equation} are set up in correspondence to the states ${\left|j_v \right\rangle}$,${\left\langle m_v \right|}$, where $v = \overline{1,N}$, 
$j = \alpha, \beta$, $m = \alpha, \beta $. For instance, the relationships for commutation rules are
\begin{equation}
\label{eq9a}
[\hat {\sigma}_v^{lm}, \hat {\sigma}_v^{pq}] = \hat {\sigma }_v^{lq} \delta_{mp} - \hat {\sigma }_v^{pm}\delta_{ql}. 
\end{equation} 
Further
\begin{equation}
\label{eq5}
\begin{split}
\raisetag{40pt}
\left[\begin{array}{*{20}c}
{\hat F^-_l}  \\
 \\
{\hat F^+_l}  \\
\\
{\hat F^z_l} 
\end{array} 
\right] = \hat {\vec F} =  \left[ {\hat {\vec {\sigma}}_l \otimes \hat {\vec {\mathcal{G}}}_{l - 1,l + 1}} \right],
\end{split}
\end{equation}
where vector operators $\hat {\vec {\mathcal{G}}}_{l - 1,l + 1}$,  $l = \overline{2,N-1}$, are given by the expressions
\begin{equation}
\label{eq6}
\hat {\vec {\mathcal{G}}}_{l - 1,l + 1} = \hat {\mathcal{G}}_{l - 1,l + 1}^-  \vec e_ +  + \hat {\mathcal{G}}_{l - 1,l + 1}^ +  \vec e_ - + \hat {\mathcal{G}}_{l - 1,l + 1}^z \vec e_z,
\end{equation}
in which 
\begin{subequations}
\label{eq7}
\begin{gather}
\hat {\mathcal{G}}_{l - 1,l + 1}^-  = -\frac{1 }{\hbar} \sum\limits_{\vec k}\hat{f}_{l \vec k} - \frac{J }{\hbar }(\hat\sigma _{l + 1}^- + \hat\sigma _{l - 1}^-) , \\
\hat {\mathcal{G}}_{l - 1,l + 1}^+ = -\frac{1}{\hbar} \sum\limits_{\vec k}\hat{f}_{l \vec k} - \frac{J}{\hbar }(\hat\sigma _{l + 1}^+ + \hat\sigma _{l - 1}^ + ), \\
\hat {\mathcal{G}}_{l - 1,l + 1}^z = - \omega_{l} - \frac{J}{\hbar }(\hat\sigma _{l + 1}^z + \hat\sigma _{l - 1}^z ).
\end{gather}
\end{subequations}
Here operator $\hat{f}_{l \vec k}$ is
 \begin{equation}
\label{eq8}
\hat{f}_{l \vec k} = v_{l \vec k} \hat{a}_{\vec k} + \hat{a}_{\vec k}^{+} {v^*}_{l \vec k}.
\end{equation}
In relations (\ref{eq7}) $J$ is the exchange interaction constant in the case of magnetic resonance transitions or its optical
analogue in the case of optical transitions, the function $v_{l \vec k}$ in (\ref{eq8}) is
\begin{equation}
\label{eq9}
 v_{l \vec k} = - \frac{1}{\hbar} p_l^{jm} (\vec e_{\vec k} \cdot \vec e_{\vec P_{l}}) \mathfrak{E}_{\vec k} e^{ - i \omega_{\vec k}t + i \vec k \vec r},
\end{equation}
where $p_l^{jm}$ is matrix element of operator of magnetic (electric) dipole moment $\vec P_{l}$ of $\textit{l-th}$ chain unit between the states $\left| {j_{l}} \right\rangle$ and 
$\left| m_{l} \right\rangle$ with $j \in \{\alpha, \beta\}$,  $m \in \{\alpha, \beta\}$, $j \neq m$, $\vec e_{\vec k}$ is unit polarization vector, $\vec e_{\vec P_{l}}$ is unit vector along $\vec P_{l}$-direction, $\mathfrak{E}_{\vec k}$ is the quantity, which has the dimension of magnetic (electric) field strength, $\vec k$ is quantized EM-field wave vector, the components of which get a discrete set of values, $\omega_{\vec k}$ is the frequency, corresponding to ${\vec k}$th mode of EM-field,  $\hat{a}^+_{\vec k}$ and $\hat{a}_{\vec k}$ are  EM-field  creation and  annihilation operators correspondingly. In the suggestion, that the contribution of spontaneous emission is relatively small, we will have $p_{l}^{jm} = p_{l}^{mj} \equiv p_{l} $, where $j \in \{\alpha, \beta\}$,  $m \in \{\alpha, \beta\}$, $j \neq m$. Further matrix $||\hat{R}^{(\lambda)}_{\vec{q}l}||$ is
\begin{equation}
\label{eq10}
\begin{split}
\raisetag{40pt}
||\hat{R}^{(\lambda)}_{\vec{q}l}|| = 
\frac{1}{i\hbar} \left[\begin{array}{*{20}c}{ 2 \hat{B}^{(\lambda)}_{\vec{q}l} \hat\sigma^-_l}  \\
 \\
{ -2 \hat{B}^{(\lambda)}_{\vec{q}l} \hat\sigma^+_l}  \\
\\
{0} \end{array} 
\right] 
\end{split}
\end{equation}
Here $\hat{B}^{(\lambda)}_{\vec{q}l}$ is
\begin{equation}
\label{eq10a}
\hat{B}^{(\lambda)}_{\vec{q}l} = \sum\limits_{\vec{q}}\lambda_{\vec{q}l} (\hat{b}^{+}_{\vec{q}} + \hat{b}_{\vec{q}}),
\end{equation} 
 $\hat{b}_{\vec q}^+$ ($\hat{b}_{\vec q}$) is the creation
(annihilation) operator of the phonon with impulse ${\vec q}$ and with
energy $\hbar \omega_{\vec q} $, $\lambda_{\vec q l}$ is electron-phonon coupling constant. In equations (\ref{eq2}) and (\ref{eq3})
$\|\sigma_P^z\| $ is Pauli $z$-matrix,  $\left\| g \right\|$ in equation (\ref{eq1}) is diagonal matrix,
numerical values of its elements are dependent on the basis choice.
It is at appropriate basis
\begin{equation}
\label{eq11}
\left\| g \right\| = 
\left[
\begin{array}{*{20}c}
 {1} & {0} & {0} \\
 {0} & {1} & {0} \\
 {0} & {0} & {1} \\
\end{array} 
\right]. 
\end{equation} 
 
Right hand side expression  in (\ref{eq5}) is vector product of vector operators.   It can be 
calculated by using of known expression (\ref{eq12}) with additional coefficient ${\frac{1}{2}}$ the only, which is appeared, since
\begin{equation}
\label{eq12}
\left[ {\hat {\vec {\sigma}} _l \otimes \hat {\vec {\mathcal{G}}}_{l - 1,l + 1} } \right] = \frac{1}{2} \left| {\begin{array}{*{20}c}
 {\vec e_- \times \vec e_z} & {\hat{\sigma}_l^-} & {\hat {\mathcal{G}}_{\,\,l - 1,l + 1}^-} \\
 {\vec e_z \times \vec e_+} & {\hat{\sigma}_l^+} & {\hat {\mathcal{G}}_{\,\,l - 1,l + 1}^+} \\
 {\vec e_+ \times \vec e_-} & {\hat{\sigma}_l^z} & {\hat {\mathcal{G}}_{\,\,l - 1,l + 1}^z} \\
\end{array}} \right|',
\end{equation}
the products of two components of two vector operators are replaced by anticommutators of corresponding 
components. Given detail is mapped by symbol $\otimes$ in (\ref{eq5}) and by symbol $'$ in determinant (\ref{eq12}). 

It follows from comparison with semiclassical Landau-Lifshitz (L-L) equation for dynamics of spectroscopic transitions in a chain of exchange coupled centers \cite{Yearchuck_Doklady}, \cite{Yearchuck_Yerchak_Dovlatova}, that the equation, which is given by (\ref{eq1}) is its QED-generalization.
 In comparison with semiclassical description, where the description of dynamics of spectroscopic 
transitions is exhausted by one vector equation (L-L equation or L-L based equation), in the case of completely quantum consideration L-L type equation describes the only one subsystem of three-part-system, which consist of EM-field, dipole moments' (magnetic or electric) matter subsystem and phonon subsystem. It was concluded in \cite{Yearchuck_Yerchak_Dovlatova}, that the presence of additional equations for description of transition dynamics by QED model in comparison with semiclassical model  leads  to a number of  new effects, which can be predicted the only by QED consideration of resonance transition phenomena. One of new effect was described in 
\cite{Yearchuck_Yerchak_Dovlatova}, starting the only from the mathematical structure of the equations. It was argued, that the equations (\ref{eq1}), (\ref{eq2}) represent
themselves vector-operator difference-differential generalization of the system, which belongs to well known family of equation systems - Volterra model systems, widely used in biological tasks of population dynamics studies, which in its turn is generalization of Verhulst equation.  In other words, it was predicted, for instance, that by some parameters in two-sybsystem Volterra model the stochastic component in solution will be appeared. Given prediction has experimental confirmation by the study of optical properties in carbynes \cite{Yearchuck_PL} indicating, that in given material strong electron-photon interaction is realized, which allows to explain the posibility to observe the stationary IR-reflection or absorbtion spectra both in usual and in stochastic regime. 

The terms like to right hand side terms in (\ref{eq3})  were used in so called "spin-boson" Hamiltonian \cite{Leggett} and in  so called "independent boson model" \cite{Mahan}. Given models were used to study phonon effects in a single quantum dot within a microcavity \cite{Heitz}, \cite{Tuerck}, \cite{Besombes},   \cite{WilsonRae}, \cite{Zhu}. So, it has been shown in \cite{WilsonRae}, \cite{Zhu}, that the presence of the term in Hamiltonian \cite{Yearchuck_Yerchak_Dovlatova}
\begin{equation}
\label{eq13}
\mathcal{\hat H}^{CPh} =\sum\limits_{j=1}^N \sum\limits_{\vec q}  \lambda_{\vec q} (\hat{b}_{\vec q}^{ +} +\hat{b}_{\vec q})\hat{\sigma}^z_j,
\end{equation} which coincides with corresponding term in Hamiltonian in \cite{WilsonRae}, \cite{Zhu} at $N = 1$ [contribution of given term to  the equations for spectroscopic transitions  is  $\pm {\sum\limits_{l = 1}^N \hat\sigma_l^{z} \lambda_{\vec q}}$, see equation (\ref{eq3}), (note that the equations for spectroscopic transitions were not derived in above cited works \cite{Heitz}, \cite{Tuerck}, \cite{Besombes},   \cite{WilsonRae}, \cite{Zhu})] leads the only to exponential decrease  of the magnitude of quantum Rabi oscillations with increase of electron-phonon coupling strength and even to their supression at relatively strong electron-phonon coupling.  However by strong electron-photon coupling amd strong electron-phonon coupling quite other picture of quantum relaxation processes becomes to be possible. Really,
if to define the 
 wave function  of the chain system, interactig
with quantized EM-field  and with quantized lattice vibration field, to be
vector of the state in Hilbert space over quaternion ring,  that is 
quaternion function of quaternion
argument, then like to \cite{Yearchuck_Yerchak_Dovlatova} can be shown, that the equations (\ref{eq1}) to (\ref{eq3}) are Lorentz invariant and the transfer to observables can be realized. In particular, taking into account, that quaternion vector of the state is proportional to spin, the   Hamiltonian, given by  (\ref{eq13}) describes in fact the interaction of phonon field  with $z$- component $S^z$ of the spin of matter subsystem. It seems to be reasonable to take into consideration the  interaction of phonon field with  $S^+$- and $S^-$ components of the spin of matter subsystem. Therefore we come in a natural way to the following  Hamiltonian
 \begin{equation}
\label{eq14}
\mathcal{\hat H} = \mathcal{\hat H}^C + \mathcal{\hat H}^F + \mathcal{\hat H}^{C F} + \mathcal{\hat H}^{Ph} + \mathcal{\hat H}^{CPh} ,
\end{equation} 
where 
${\mathcal{\hat H}^C}$ is chain Hamiltonian by the absence of the interaction with EM-field, ${\mathcal{\hat H}^F}$ is field Hamiltonian, ${\mathcal{\hat H}^{C F}}$ is Hamiltonian, describing the interaction between quantized EM-field and atomic chain.
 Hamiltonian ${\mathcal{\hat H}^C}$ is
\begin{equation}
\label{eq15}
{\mathcal{\hat H}^C} = {\mathcal{\hat H}^0} + {\mathcal{\hat H}^J}, 
\end{equation} 
where ${\mathcal{\hat H}^0}$ is chain Hamiltonian in the absence of the interaction between structural elementary units of the chain.
  ${\mathcal{\hat H}^0}$ is given by the expression
\begin{equation}
\label{eq16}
\mathcal{\hat H}^0 = \sum\limits_{v=1}^N \sum\limits_m E_{mv}{\left|m_v \right\rangle} {\left\langle m_v \right|}.
\end{equation}
Here $m = \alpha, \beta$, $E_{mv}$ are eigenvalues of ${\mathcal{\hat H}^0}$, which correspond to the states ${\left|m_v \right\rangle}$ of $v\textit{th}$ chain unit.                   
Hamiltonian ${\mathcal{\hat H}^J}$ is
\begin{equation}
\label{eq17}
\begin{split}
\raisetag{40pt}
 \mathcal{\hat H}^J = 
 \sum\limits_{n = 1}^N [J_{_{E}} (\hat {\sigma} _n^ + \hat {\sigma} _{n + 1}^- + \hat {\sigma} _n^ -  \hat {\sigma} _{n + 1}^ + + \frac{1}{2}\hat {\sigma} _n^z \hat {\sigma} _{n + 1}^z ) + H.c.].
\end{split}
\end{equation}
It is suggested 
in the model, that $\left| {\alpha _n } \right\rangle $ and $\left| {\beta _n } \right\rangle $ are eigenstates, producing the full set for each of $N$ elements. It is evident, that given assumption can be realized strictly the 
only by the absence of the interaction between the elements. At the same 
time proposed model will rather well describe the real case, if the 
interaction energy of adjacent elements is much less of the energy of the 
splitting $\hbar \omega _0 =\mathcal{E}_\beta -\mathcal{E}_\alpha$ between the energy levels, 
corresponding to the states $\left|\alpha_n\right\rangle$ and $\left|\beta_n\right\rangle$. The case considered includes in fact all known 
experimental situations.
 It is clear, that Hamiltonian $\mathcal{\hat H}^{C F}$ of interaction of quantized EM-field with atomic chains can also be represented in the set of variables, which includes the components of spectroscopic transition vector operator $\hat {\vec {\sigma }}_v$. Really, suggesting dipole approximation to be true and polarization of field components to be fixed, we have 
\begin{equation}
\label{eq18}
\begin{split}
\raisetag{40pt}
\mathcal{\hat H}^{C F} = -\sum\limits_{j = 1}^n \sum\limits_{l \neq m} \sum\limits_{m} \sum\limits_{\vec k} [ p_j^{lm} \hat {\sigma}_j^{lm} (\vec e_{\vec k} \vec e_ {\vec P_j}) \mathfrak{E}_{\vec k} \hat{a}_{\vec k} \times \\
e^{ - i \omega_{\vec k} t+ i \vec k \vec r} + H.c. ], 
\end{split}
\end{equation}
where $p_j^{lm}$ is matrix element of operator of magnetic (electric) dipole moment $\vec P_j$ of $\textit{j-th}$ chain unit between the states $\left| {l_j} \right\rangle$ and 
$\left| {m_j} \right\rangle$ with $l_j = \alpha_j, \beta_j, m_j = \alpha_j, \beta_j$, $\vec e_{\vec k}$ is unit polarization vector, $\vec e_{\vec P_j}$ is unit vector along $\vec P_j$-direction, $\mathfrak{E}_{\vec k}$ is the quantity, which has the dimension of  magnetic (electric) field strength, $\vec k$ is wave vector, $\hat{a}_{\vec k}$ is field annihilation operator. In the suggestion, that the contribution of spontaneous emission is relatively small, we will have $p_j^{lm} = p_j^{ml} \equiv p_j $, where $l = \alpha, \beta, m = \alpha, \beta$. Let define the function
\begin{equation}
\label{eq19}
 q_{j \vec k} = - \frac{1}{\hbar} p_j (\vec e_{\vec k} \cdot \vec e_{\vec P_j}) \mathfrak{E}_{\vec k} e^{ - i \omega_{\vec k} t+ i \vec k \vec r}
\end{equation}
Then the expression (\ref{eq18}) can be rewritten in the form
\begin{equation}
\label{eq20}
\mathcal{\hat H}^{C F} = \sum\limits_{v = 1}^n \sum\limits_{\vec k} [q_{j \vec k} (\hat {\sigma}_j^- + \hat {\sigma}_j^+) \hat{a}_{\vec k} + (\hat {\sigma}_j^- + \hat {\sigma}_j^+) \hat{a}_{\vec k}^{ +} {q^*}_{j \vec k}],
\end{equation}
where $\hat{a}_{\vec k}^{ +}$ is EM-field creation operator, $\hat{a}_{\vec k}^{ }$ is EM-field annihilation operator, superscript $^*$ in ${q^*}_{j \vec k}$ means complex conjugation.
 Field Hamiltonians have usual form
\begin{equation}
\label{eq21}
\mathcal{\hat H}^{F} = \sum\limits_{\vec k} \hbar \omega_{\vec k} (\hat{a}_{\vec k}^{ +} \hat{a}_{\vec k} + \frac{1}{2})
\end{equation}
for EM-field and
\begin{equation}
\label{eq22}
\mathcal{\hat H}^{Ph} = \sum\limits_{\vec q} \hbar \omega_{\vec q} (\hat{b}_{\vec q}^{ +} \hat{b}_{\vec q} + \frac{1}{2})
\end{equation} for phonon field.
 Hamiltonian $\mathcal{\hat H}^{CPh}$ is
\begin{equation}
\label{eq23}
\mathcal{\hat H}^{CPh} = \mathcal{\hat H}^{CPh}_z + \mathcal{\hat H}^{CPh}_\pm, 
\end{equation} 
where $\mathcal{\hat H}^{CPh}_z$ is determined by the expression

\begin{equation}
\label{eq24}
\mathcal{\hat H}^{CPh}_z =\sum\limits_{j=1}^N \sum\limits_{\vec q} (\lambda^z_{\vec q} \hat{b}_{\vec q} + (\lambda^z_{\vec q})^*\hat{b}_{\vec q}^{+})\hat{\sigma}^z_j.
\end{equation}
 Hamiltonian $\mathcal{\hat H}^{CPh}_\pm$ can be represented in the following form
\begin{equation}
\label{eq25}
\mathcal{\hat H}^{CPh}_\pm = \sum\limits_{j=1}^N \sum\limits_{\vec q}\lambda^\pm_{\vec q} (\hat{\sigma}^-_j + \hat{\sigma}^+_j) \hat{b}_{\vec q} + (\lambda^\pm_{\vec q})^* (\hat{\sigma}^-_j + \hat{\sigma}^+_j)\hat{b}_{\vec q}^{+}.
\end{equation}
Here  $\lambda^z_{\vec q}$ and $\lambda^\pm_{\vec q}$ are  electron-phonon coupling constants, which characterisire correspondingly the interaction with $z$- component $S^z_j$ and  with  $S^+$- and $S^-_j$ components of the spin of jth chain unit. It seems to be understandable, that they can be different in general case. Moreover, in order to take into account the interaction with both equilibrium and nonequilibrium phonons  both the electron-phonon coupling constants have to be complex numbers, that takes proper account by expressions (\ref{eq24}), (\ref{eq25}). 

It can be shown, that
the equations of the motion for spectroscopic transition operators $\hat {\vec {\sigma }}_l$,  for quantized 
 EM-field operators $\hat{a}_{\vec k}$, $\hat{a}_{\vec k}^{ +}$ and for phonon field operators $\hat{b}_{\vec q}$, $\hat{b}_{\vec q}^{ +}$  are the following. Instead equation (\ref{eq1}) we have
\begin{equation}
\label{eq26}
\begin{split}
\raisetag{40pt}\frac{\partial}{\partial t} \left[\begin{array}{*{20}c}
{\hat\sigma^-_l}  \\
 \\
{\hat\sigma^+_l}  \\
\\
{\hat\sigma^z_l} 
\end{array} 
\right] = 2 \left\| g \right\| \left[\begin{array}{*{20}c}
{\hat F^-_l}  \\
 \\
{\hat F^+_l}  \\
\\
{\hat F^z_l} 
\end{array} 
\right] + ||\hat{R}^{(\lambda^z)}_{\vec{q}l}|| + ||\hat{R}^{(\lambda^\pm)}_{\vec{q}l}||, 
\end{split}
\end{equation}
where matrix $||\hat{R}^{(\lambda^z)}_{\vec{q}l}||$ is
\begin{equation}
\label{eq27}
\begin{split}
\raisetag{40pt}
||\hat{R}^{(\lambda^z)}_{\vec{q}l}|| = 
\frac{1}{i\hbar} \left[\begin{array}{*{20}c}{ 2 \hat{B}^{(\lambda^z)}_{\vec{q}l} \hat\sigma^-_l}  \\
 \\
{ -2 \hat{B}^{(\lambda^z)}_{\vec{q}l} \hat\sigma^+_l}  \\
\\
{0} \end{array} 
\right] 
\end{split}
\end{equation}
with  $\hat{B}^{(\lambda^z)}_{\vec{q}l}$, which is given by
\begin{equation}
\label{eq28}
\hat{B}^{(\lambda^z)}_{\vec{q}l} = \sum\limits_{\vec{q}}[(\lambda^z_{\vec{q}l})^* \hat{b}^{+}_{\vec{q}} + \lambda^z_{\vec{q}l} \hat{b}_{\vec{q}}].\end{equation} 
Matrix $||\hat{R}^{(\lambda^\pm)}_{\vec{q}l}||$ is
\begin{equation}
\label{eq29}
\begin{split}
\raisetag{40pt}
||\hat{R}^{(\lambda^z)}_{\vec{q}l}|| = 
\frac{1}{i\hbar} \left[\begin{array}{*{20}c}{ -\hat{B}^{(\lambda^\pm)}_{\vec{q}l} \hat\sigma^z_l}  \\
 \\
{  \hat{B}^{(\lambda^\pm)}_{\vec{q}l} \hat\sigma^z_l}  \\
\\
{\hat{B}^{(\lambda^\pm)}_{\vec{q}l} (\hat\sigma^+_l - \hat\sigma^-_l)} \end{array} 
\right],  
\end{split}
\end{equation}
where $\hat{B}^{(\lambda^\pm)}_{\vec{q}l}$ is
\begin{equation}
\label{eq30}
\hat{B}^{(\lambda^\pm)}_{\vec{q}l} = \sum\limits_{\vec{q}}[(\lambda^\pm_{\vec{q}l})^* \hat{b}^{+}_{\vec{q}} + \lambda^\pm_{\vec{q}l} \hat{b}_{\vec{q}}].\end{equation} 
The equation (\ref{eq2}) remains without changes. The equation (\ref{eq3}) is
\begin{equation}
\label{eq30a}
\begin{split}
\raisetag{40pt}
&\frac{\partial}{\partial t} 
\left[
\begin{array}{*{20}c}
 {\hat{b}_{\vec k^{}}} \\
 \\
 {\hat{b}_{\vec q^{}}^+} \\
\end{array} 
\right] = -i \omega_{\vec q^{}} ||\sigma_P^z|| \left[\begin{array}{*{20}c}
 {\hat{b}_{\vec q^{}}} \\
 \\ 
 {\hat{b}_{\vec q^{}}^+} \\
\end{array} 
\right] 
 + \\
&\frac{i}{\hbar}
\left[
\begin{array}{*{20}c}
{-\sum\limits_{l = 1}^N \{\lambda^z_{\vec q l} \hat\sigma_l^{z} + \lambda^{\pm}_{\vec q l} (\hat\sigma_l^{+} + \hat\sigma_l^{-})\}} \\
\\
{\sum\limits_{l = 1}^N \{\lambda^z_{\vec q l} \hat\sigma_l^{z} + \lambda^{\pm}_{\vec q l} (\hat\sigma_l^{+} + \hat\sigma_l^{-})\}} \\
\end{array} \right].
\end{split}
\end{equation}

 Thus, QED - QPFT model for  dynamics of spectroscopic transitions in 1D multiqubit  exchange coupled  system  is generalized by taking into account the earlier proof \cite{Yearchuck_Yerchak_Dovlatova}, that spin vector is  quaternion vector of the state of any quantum systen in Hilbert space defined  over quaternion ring and consequently all the spin components has to be taken into account. New quantum  phenomenon is predicted. The prediction results from the structure of  the equations derived and it consists in the following. The coherent system of the resonance phonons, that is,  the phonons with the energy, egualed to resonance photon  energy can be formed by resonance, that can lead to appearance along with   Rabi oscillations determined by spin (electron)-photon coupling with the frequency $\Omega^{RF}$ of Rabi oscillations determined by spin (electron)-phonon coupling with the frequency $\Omega^{RPh}$. In other words QED - QPFT model predicts the oscillation character of quantum relaxation, that is quite different character in comparison with phenomenological and semiclassical Bloch models. Moreover if $\mid\lambda^{\pm}_{\vec q l}\mid < g$ the second Rabi oscillation process will be observed by stationary state of two subsystems \{ EM-fied + magnetic (electric) dipoles \}, that is, it will be registered in quadrature with the first Rabi oscillation process.  It can be experimentally detected even by stationary spectroscopy methods.

 \subsection{Experiment} 
ESR studies in anthracite samples of medium-scale metamorphism have been carried out on stationary ESR spectrometer "Radiopan" at room temperature. The  100 kHz high frequency modulation of static magnetic field $H_0$ was used. The static magnetic field $H_0$ was sweeped in two directions from lesser value to bigger  value ($\frac{dH_0}{dt} > 0$) and from bigger value to lesser value ($\frac{dH_0}{dt} < 0$). Two regime of registration were used - with automatic microwave frequency adjustment (AF-mode) and without automatic microwave frequency adjustment with constant microwave frequency (CF-mode). The spectra obtained are presented in Figures 1, 3-9. The most interesting, that the spectra can be registered by using of 100 kHz high frequency (HF) modulation canal at modulation amplitude $H_m = 0$, Figure 1. It is seen from Figure 1, that the spectrum has very unusual for stationary ESR-spectroscopy form, it is similar partly to the spectra, registered by nonstationary transient ESR-spectroscopy. At the same time the functional dependence is another. The spectral distribution of nutation signals in  transient ESR-spectroscopy is described usually by continuous Bessel functions. The spectrum, presented in  Figure 1, has time-discrete character and  consist of two group of very narrow ESR lines. For instance, the linewidth ${\Delta}H$ of the central line with maximal amplitude in left group is equal to $0.01 (\pm 0.004)$ G. It allows to determine $g$-value (that is g-factor in Zeeman term of spin-Hamiltonian) very precisely. It is equal to 2.002735  $(\pm 0.000001)$. The linewidth of the other lines is practically the same. 
The lines presented in  Figure 1 seem to be the most narrow among all the lines, registered in ESR-spectroscopy at all. The position of the central line with maximal amplitude in right group is given by 
$g$-value 2.00255 $(\pm 0.00001)$. The structure of the spectrum is very similar to  theoretical time evolution for a system with one qubit in JCM, Figure 2. Really, the spectra in both the groups can be described by almost sinusoidal dependence  with asymmetric Gaussian envelope, Figure 3. The qualitative difference between time evolution given by JCM and spectral distribution  presented in  Figure 1 is that, that the left group of lines in JCM has cosinusoidal dependence, Figure 2. The possibility to explain the spectrum observed by usual amplitude modulation from possible technical noise (determined for instance by the operation of static magnetic field stabilization unit or microwave frequency stabilization unit)  is excluded, since the envelope by amplitude modulation is described by sinusoidal dependence, we have the envelope, described by Gauss distribution with slightly different parameters for left and right sides. So the fit of experimental data, Figure 3, for left hand side group of lines is given by the following expression
\begin{equation}
\label{eq38}
y = y_0 + \frac{A}{W} \sqrt{\frac{2}{\pi}} \exp\{{-2\left[\frac{(H - H_c)}{W}\right]^2}\},
\end{equation} 
where $y_0 = 52.4$, $H_c = 3322.95 G$, $W = 0.36 G$, $A = 32 G$.

\begin{figure}
\includegraphics[width=0.5\textwidth]{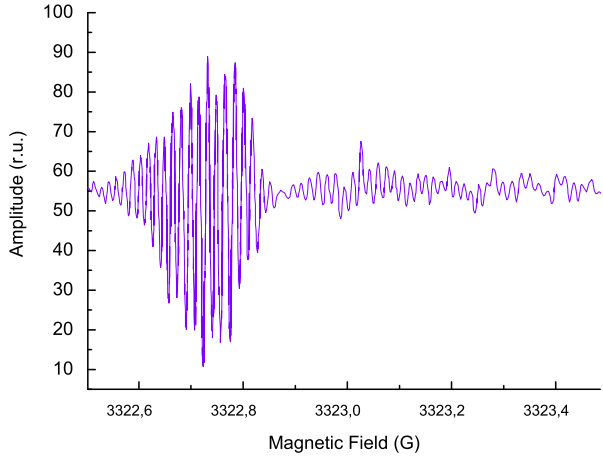}
\caption[Quantum Rabi oscillation process indicating on the formation of the coherent system of the resonance phonons$]
{\label{Figure1} Quantum Rabi oscillation process indicating on the formation of the coherent system of the resonance phonons}
\end{figure}

\begin{figure}
\includegraphics[width=0.47\textwidth]{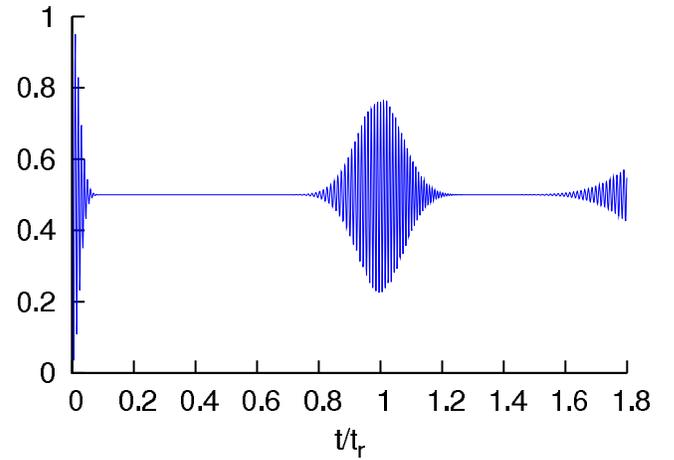}
\caption[Theoretical evolution for  JCM-system, initial state $|\psi\rangle \otimes |\alpha\rangle$ of which is direct product of atomic ground state and coherent field state with $\overline{n} = 50$]
{\label{Figure2} Theoretical evolution for  JCM-system, initial state $|\psi\rangle \otimes |\alpha\rangle$ of which is direct product of atomic ground state and coherent field state with $\overline{n} = 50$}
\end{figure}

\begin{figure}
\includegraphics[width=0.5\textwidth]{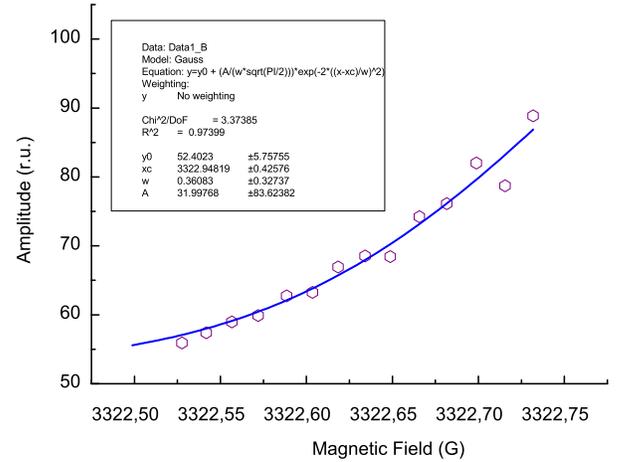}
\caption[The fit of left hand side envelope of left oscillation group in the spectrum, presented in Figure 1 by Gaussian]
{\label{Figure3} The fit of left hand side envelope of left oscillation group in the spectrum, presented in Figure 1 by Gaussian}\end{figure}

\begin{figure}
\includegraphics[width=0.5\textwidth]{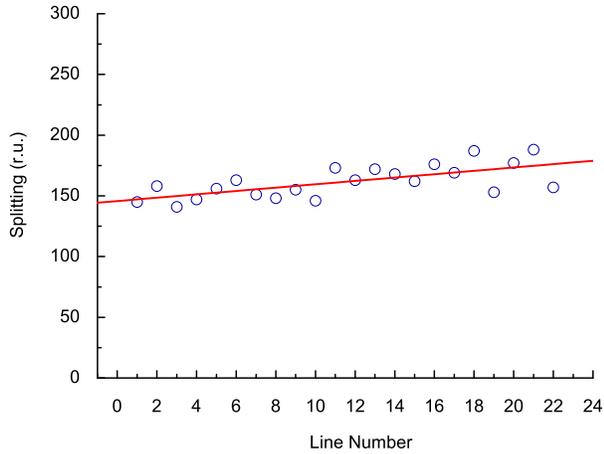}
\caption[The fit of the splitting between the lines in left oscillation group in the spectrum, presented in Figure 1 by linear dependence]
{\label{Figure4} The fit of the splitting between the lines in left oscillation group in the spectrum, presented in Figure 1 by linear dependence}\end{figure}
\begin{figure}
\includegraphics[width=0.5\textwidth]{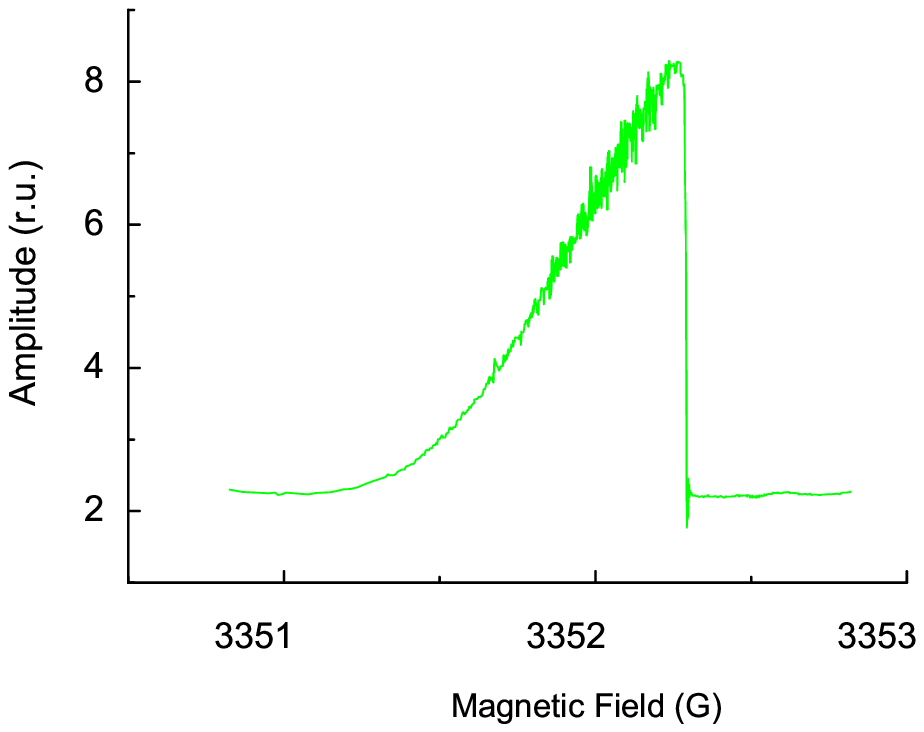}
\caption[ESR spectrun of anthracite sample by automatic microwave frequency adjustment, $\frac {dH_0}{dt} > 0$]
{\label{Figure5} ESR spectrun of anthracite sample by automatic microwave frequency adjustment, $\frac {dH_0}{dt} > 0$}\end{figure}
\begin{figure}
\includegraphics[width=0.5\textwidth]{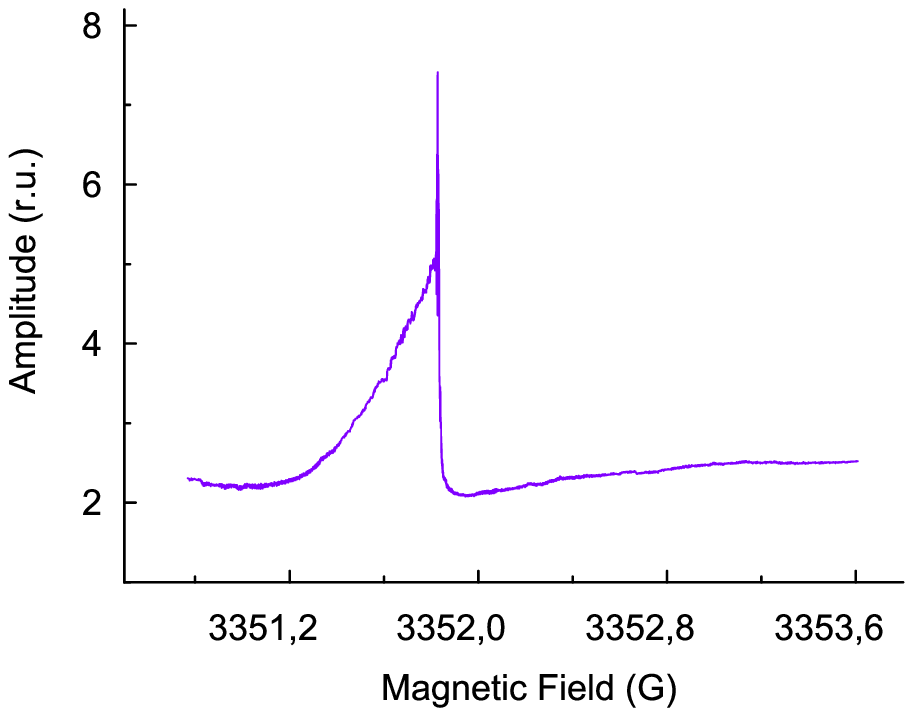}
\caption[ESR spectrun of anthracite sample by automatic microwave frequency adjustment, $\frac {dH_0}{dt} < 0$]
{\label{Figure6} ESR spectrun of anthracite sample by automatic microwave frequency adjustment, $\frac {dH_0}{dt} < 0$}\end{figure}
\begin{figure}
\includegraphics[width=0.5\textwidth]{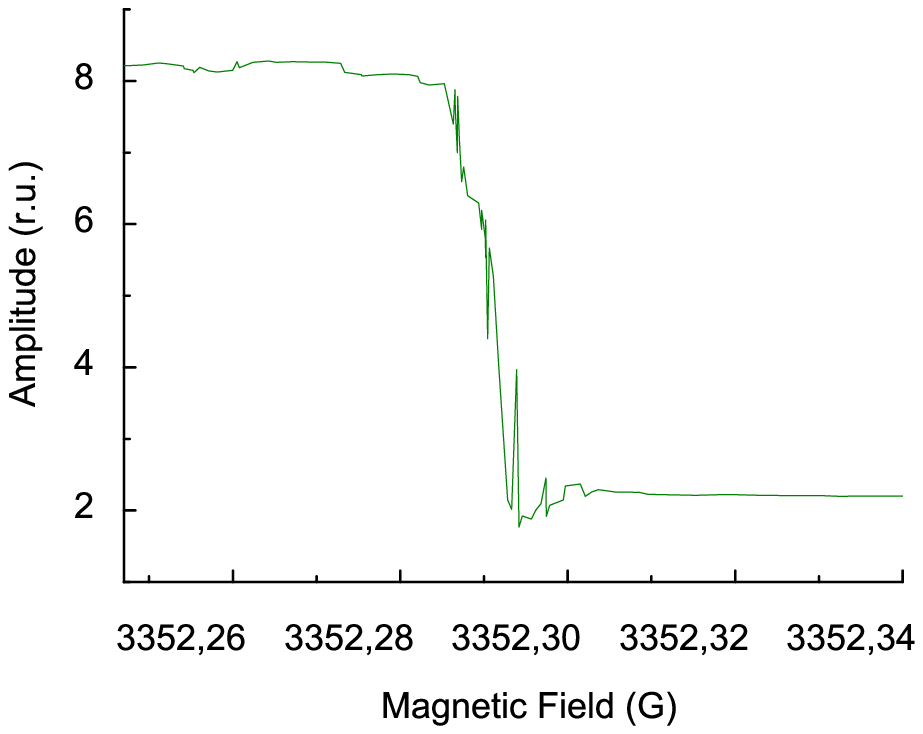}
\caption[Central part of ESR spectrun of anthracite sample with automatic microwave frequency adjustment, $\frac {dH_0}{dt} > 0$]
{\label{Figure7} Central part of ESR spectrun of anthracite sample with automatic microwave frequency adjustment, $\frac {dH_0}{dt} > 0$}\end{figure}

\begin{figure}
\includegraphics[width=0.5\textwidth]{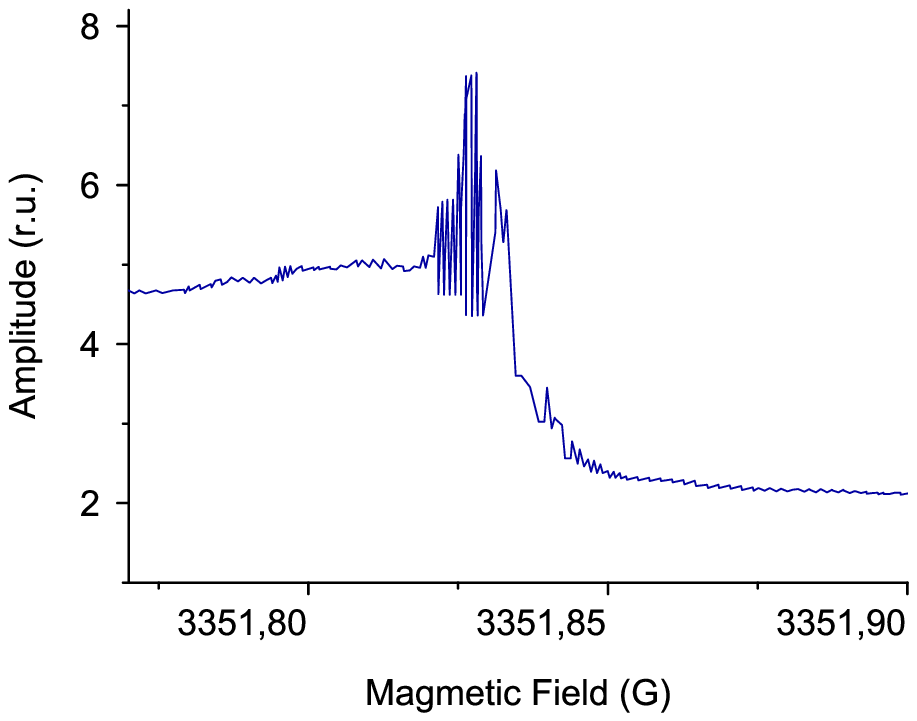}
\caption[Central part of ESR spectrun of anthracite sample with automatic microwave frequency adjustment, $\frac {dH_0}{dt} < 0$]
{\label{Figure8} Central part of ESR spectrun of anthracite sample with automatic microwave frequency adjustment, $\frac {dH_0}{dt} < 0$}\end{figure}

\begin{figure}
\includegraphics[width=0.5\textwidth]{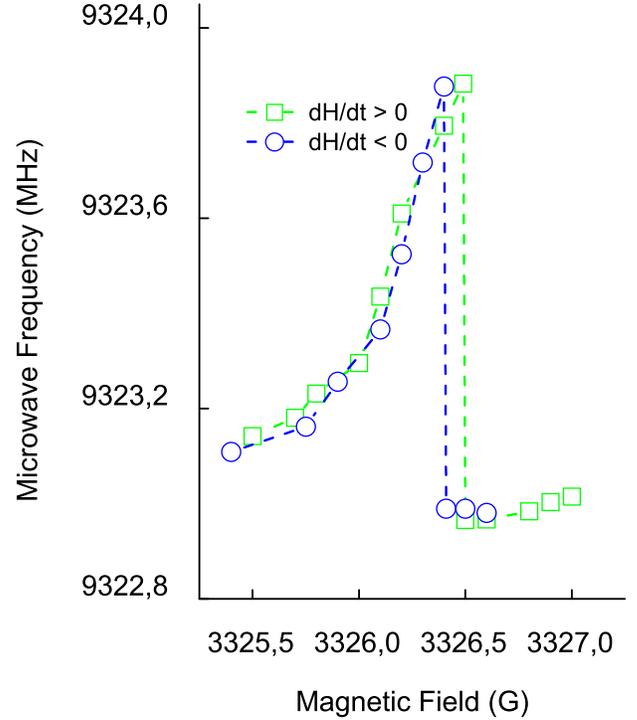}
\caption[Dependence of the frequency of measuring cavity by the ESR absorption ($\frac {dH_0}{dt} > 0$) and emission ($\frac {dH_0}{dt} < 0$) spectrum registration of anthracite sample   with automatic microwave frequency adjustment]
{\label{Figure9}Dependence of the frequency of measuring cavity by the ESR absorption ($\frac {dH_0}{dt} > 0$) and emission ($\frac {dH_0}{dt} < 0$) spectrum registration of anthracite sample   with automatic microwave frequency adjustment}\end{figure}

\begin{figure}
\includegraphics[width=0.5\textwidth]{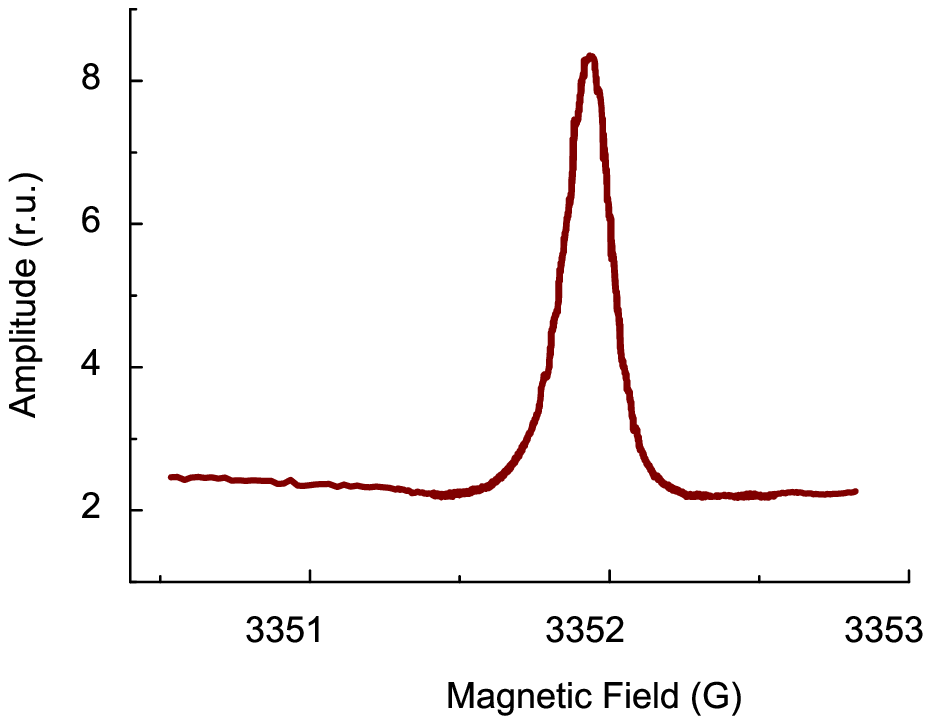}
\caption[ESR spectrun of anthracite sample without automatic microwave frequency adjustment, $\frac {dH_0}{dt} > 0$]
{\label{Figure10}ESR spectrun of anthracite sample without automatic microwave frequency adjustment, $\frac {dH_0}{dt} > 0$}\end{figure}

\begin{figure}
\includegraphics[width=0.5\textwidth]{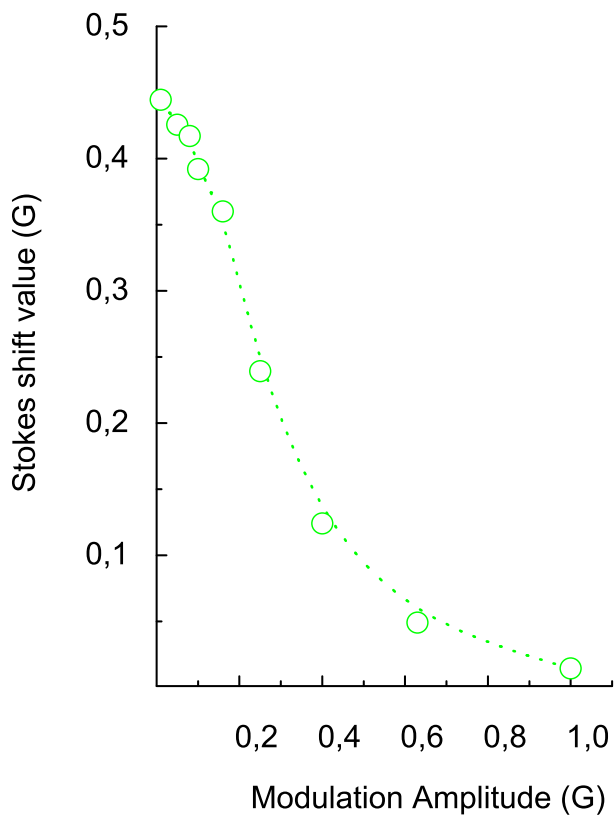}
\caption[Dependence of the Stokes shift value in ESR-ASR spectra of anthracite sample on the amplitude of 100 kHz high frequency modulation of static magnetic field]
{\label{Figure11}Dependence of the Stokes shift value in ESR-ASR spectra of anthracite sample on the amplitude of 100 kHz high frequency modulation of static magnetic field}\end{figure}

\begin{figure}
\includegraphics[width=0.5\textwidth]{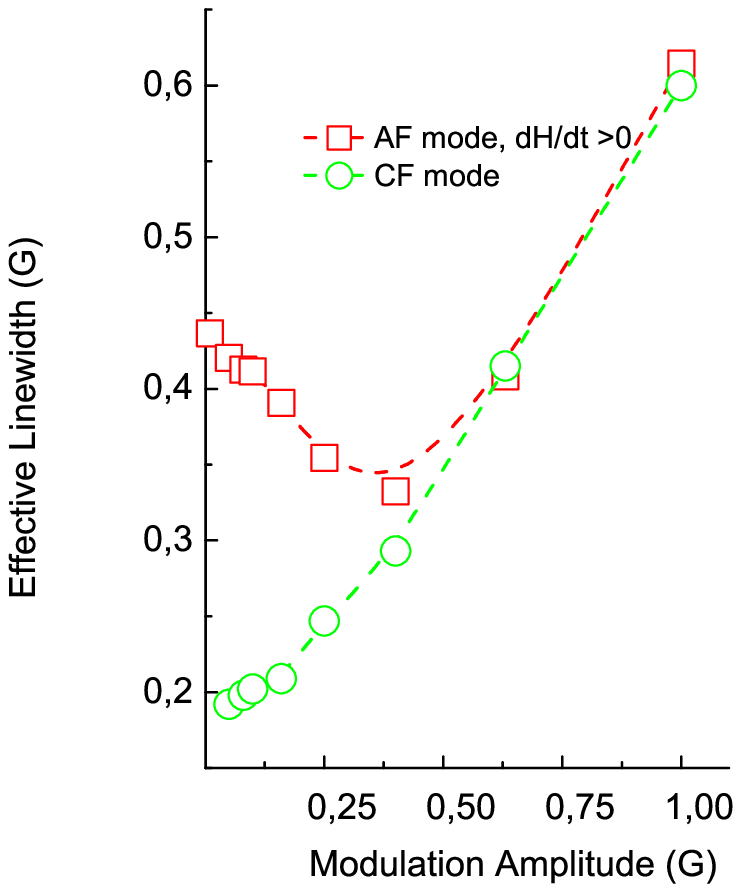}
\caption[Dependence of the effective linewidth in ESR spectra of anthracite sample on the amplitude of  high frequency modulation of static magnetic field for both AF and CF registration modes]
{\label{Figure12}Dependence of the effective linewidth in ESR spectra of anthracite sample on the amplitude of  high frequency modulation of static magnetic field for both AF and CF registration modes}\end{figure}

Further the value of g-factor of the center of the first group agrees well with g-factor of ESR line rof the same sample registered by usual conditions with nonzero amplitude of 100 kHz high frequency modulation [however it was determined substantionally more precisely]. The dependence of the amplitude and the shape of the first-group-lines and dependence of the amplitude, the shape and the position of the second-group-lines ( which are not analysed in given communication on the microwave power) is addditional argument for reality of derection of spin system response at $H_m = 0$.
Given arguments and comparison with Figure 2 allow to conclude that the spectrum in Figure 1 represents itself the quantum Rabi oscillation picture. Really, let us touch on JCM in some details. Central role in JCM plays the sum with infinite summation limit, which represents on a scale [-1,1] the degree of excitation of two-level system resulting of interaction between the atom dipole or spin and single mode quantized EM-field. It is
\begin{equation}
\label{eq31}
\begin{split}
\langle\hat{\sigma}^{z}(t)\rangle = -\exp[-|\alpha|^2]\sum_{n=1}^{\infty}\frac{|\alpha|^{2n}}{n!} \cos{2g\sqrt{\overline{n}}t},
\end{split}
\end{equation} 
where $|\alpha\rangle$ is fully coherent state of the field, taking place at $t = 0$, at that $|\alpha|^2$ = $\overline{n}$, that is, it is  the average number of photons
in the field, $g$ is coupling constant between field  and atom (spin). The sum in  (\ref{eq31}) cannot be expressed  exactly in analytical form. For very short times and very large $\overline{n}$ behavior of $\langle\hat{\sigma}^{z}(t)\rangle$ is determined by $\cos{2g\sqrt{\overline{n}}t}$.
Cummings \cite{Cummings} has shown, that by resonance and for intermediate time $t$ values the cosine Rabi oscillation damp quickly (so called collapse takes place). Given damping can be described by Gaussian envelope
\begin{equation}
\label{eq32}
\exp[-\frac{1}{2}(gt)^2].
\end{equation}
 It is substantial, that it  not depends on field intensity unlike to semiclassical Rabi oscillation damping process and it is determined entirely  the only by coupling constant $g$. The authors of the work \cite{Eberly} have found, that JCM contains so called revival process with revival time $T_R$, given by the expression
\begin{equation}
\label{eq33}
T_R = \frac{\pi}{g^2} \sqrt{\Delta^2 + 4 g^2 \overline{n}}, 
\end{equation}
where $\Delta$ is deviation of field mode frequency from resonance value. Revival process takes place at all time values, satisfying the relation $t$ = $k T_R$, $k \in N $. It is seen from  (\ref{eq33}) that revival time depends on $\overline{n}$ and it is proportional to field amplitude at $\Delta = 0$ like to dependence of Rabi frequency by semiclassical consideration. Let us illustrate the dynamics of spectroscopic transitions in JCM. The initial state for single qubit system can be defined  by direct product of the two level  matter (atomic, spin and so on) subsystem state $|\psi\rangle$ and coherent EM-field state $|\alpha\rangle$. It is the following
\begin{equation}
\label{eq34}
|\Psi(0)\rangle = |\psi\rangle \otimes |\alpha\rangle,
\end{equation}
where $|\psi\rangle$ is 
\begin{equation}
\label{eq35}
 |\psi\rangle = c_1 |\psi_1\rangle + c_2 |\psi_2\rangle \end{equation}
and $|\alpha\rangle$ is 
\begin{equation}
\label{eq36}
|\alpha\rangle =
\exp[\frac{-|\alpha|^2}{2}]\sum_{n=1}^{\infty}\frac{|\alpha|^{n}}{n!}|n\rangle, \alpha = \sqrt{\overline{n}} e^{i\phi}.
\end{equation}
 The Rabi oscillations of the probability, that the qubit is in the initial state in the first damping stage (collapse), on
a time scale of $t_c \simeq \frac{\sqrt{2}}{g}$ and then revive at $T_R \simeq  \frac{2\pi\sqrt{\overline{n}}}{g}$ are  illustrated in Figure 2 for $c_1 = 1$, $ c_2 = 0$ by plotting 
\begin{equation}
\label{eq37}
\sum_{n=1}^{\infty}|\langle\psi_1, n|\Psi(t)\rangle |^2,
\end{equation}
where $\langle\psi_1, n| = \langle\psi_1|\otimes \langle n|$  is the ground  state for the
qubit  with $n$ photons in the cavity. Our measurements of the dependence of the ESR spectral characteristics on the microwave power (photon number) is also showed that in correspondence with JCM the positions of the lines of left hand side group are not dependent on photon number, at the same time the positions of the lines of right hand side group are  dependent on photon number.
The aforesaid difference between our results and JCM can be easily explained if we determine the Rabi frequency value. It is equaled to 28 kHz. Given value is substantially lesser in the same microwave power range, than the Rabi frequency values in excited by microwave power systems. So, for instance, in Si the Rabi frequency changes from $\approx 1 \times 10^6 rad s^{-1}$ to $\approx 4.5 \times 10^6 rad s^{-1}$ for the Si-P3 centers \cite{Yerchak_Stelmakh}. It means that we really have observed the quantum relaxation oscillations, predicted theoretically in Section 2A. The prediction of sinusoidal character (instead cosinusoidal in JCM) is also confirmed. The comparison with the model for dynamic of spectroscopic transitions, given in  Section 2A is correct, since at modulation amplitude $H_m = 0$ the absorption cannot be registered. At the same time the emission can be registered, since by detection  at 100 kHz the quantity which changes essentially slowly is registered like to usual ESR detection. Then, if to omit the part from Hamiltonian, corresponding to EM-field and z-relaxation term and to set N equaled to 1, we will have in rotating wave approximation  JCM Hamiltonian for phonon subsystem. The linear dependence of splitting, presented in Figure 4, can be explained by the  dependence of $\Omega^{RPh}$ on the static H-field, since by the change of static H-field the distance between the energy levels in qubits is also changes. Expanding  the  dependence $\Omega^{RPh}(H)$ in Taylor series and restricting by linear term (it is correct, since H-change is small), we obtain the agreement with experiment. 

 To confirm the conclusion on coherent emission character (which is like to maser emission character) of the spectrum, observed by $H_m = 0$ and to  establish the structure of the centers, which can emit stationary, we have undertaken the study of ESR-responce at various values of HF-modulation amplitude. The spectra, registered by 100 kHz HF-modulation of static magnetic field with amplitude $H_m = 0.01 G$ are presented in Figures 5 to 9. The lines have also unusual shape  for absorption derivative. The spectra are also very different by the change of sweep direction, compare Figures 5 and 6, 7 and 8, which all were registered  with automatic microwave frequency adjustment. It is seen from Figures 5 and 6, that for both static field sweep directions one of the spectrum wings has stepwise shape, which usually is characteristic by AF adjustment cycle skip. At the same time cycle skip does not takes place, that is confirmed by detailed registration of central part of the spectra in Figures 5 and 6, see Figures  7 and 8, and by simultaneous (parallel) registration of dependence of the frequency of measuring cavity, that is time dispersion registration, Figure 9. It is seen from Figure 9, that microwave frequency jumps are not exceeding 1 MHz, that is, they are on the order of value less than the adjustment range of AF adjustment unit, which is $\simeq$ 10 MHz. It is seen also  from Figure 9, that time dispersion responce has practically the same amplitude for the both sweep directions in distinction from the spetra, presented in Figures 5 and 6, where the amplitude and intensity of resonance responce at $\frac {dH_0}{dt} > 0$ markendly exceed the amplitude and the intensity of  resonance responce at $\frac {dH_0}{dt} < 0$. So, amplitude ratio is equal 2, intensity ratio is evaluated to be equal $\approx 4.6$. It is interesting, that the step position is different for $\frac {dH_0}{dt} > 0$ and $\frac {dH_0}{dt} < 0$, that is original hysteresis is taking place. All the  differences for both sweep directions disappear by the registration  without automatic microwave frequency adjustment, and for both sweep directions we obtain the spectrum, presented in Figure 10. At the same time the line  has Dyson shape \cite{Dyson}, which is characteristic for ESR absorbtion in metals and it is determined by the space dispersion contribution \cite{Erchak_Zaitsev_Stel'makh}, which is appeared  in conductive media. Dispersionlike shape of absorption ESR-response  is also the indication that the ESR centers are also strongly mobile [in the case of static PC in conductive media the space dispersion contribution and absorption contribution to resulting ESR responce is 1 to 1 \cite{Erchak_Zaitsev_Stel'makh}]. On Dyson effect with the strong mobility (strong spin diffusion) of  PC indicates  the evident deviation of line shape from Lorentz or Gauss usual dispersion line shapes.  Really. experimental value of asymmetry extent $\frac{A}{B}$ is equal to $23.4 \pm 0.5$. (Compare with $\frac{A}{B}$ = 8 and $\frac{A}{B}$ = 3.5 for usual Lorentz and Gauss dispersion line shapes correspondingly). It means in its turn, that in anthracite samples the electron-(spin)-photon interaction is really strong for both resonance and nonresonance interations. In result at relatively large amount of PC (necessary condition, which was formulated earlier) the sufficient condition for formation of the second cavity in measuring cavity is realised. It is the following. The microwave power, that is microwave photons having occurred in the sample remain [are captured and entirely absorbed] in the sample like to microwave power in measuring cavity with metal walls. Any transmission of microwave power through the samplr does not take place. It is the sufficient condition for the formation of the second cavity in measuring cavity. In other words the sample itself becomes to be the second cavity, that clearly was observed to be additional narrow peak on microwave generator generation zone oscillogramme. By AF-mode registration the microwave generator frequency follows  the changes  in the frequency of the sample-cavity, determined by time dispersion of the sample-cavity. The changes in  frequency of the order of 1 MHz are relatively large and the resonance phonons with starting frequency remain in coherent state in the sample [they cannot interact  immediately with microwave photons, which have now the other energy value]. In fact lifetime of coherent state of resonance phonons is determined by their interaction with nonresonance phonons and it can be very long, if given interaction is relatively weak. It takes place in the anthracite samples studied and it seems to be the consequence of their disordered structure. By the sweep back of the  static magnetic field $H_0$ ($\frac{dH_0}{dt} < 0$) the well known phenomenon of acoustic spin resonance (ASR) is realised, which is characterised by own Rabi oscillation process, presented in Figure 1. It is especially interesting, that acoustic Rabi oscillation process has quantum nature with discrete dependence on the time (time sweep is connected with 
the  static magnetic field sweep). The possibility to register the  ASR by means of 100-kHz modulation canal is determined by the following factors. AF unit produces frequency modulation of microwave power, which becomes to be not srongly monochromatic. At the same time the quantum Rabi oscillation process itself is not  monochromatic process. Therefore the nonmonochromatic modulation of the  static magnetic field is taking place, and the Fourier component of given  nonmonochromatic modulation is registered by 100-kHz modulation canal. Given conclusion is confirmed by the polarity of derivative of ASR responce. Really,
it is substantial, that observed polarity of derivative of ESR-ASR responce by usual registration in AF-mode  does not depend on sweep direction. At the same time, when ESR spectra by $\frac{dH_0}{dt} > 0$ and by $\frac{dH_0}{dt} < 0$ are tuned for the absorbtion registration the polarities of qiven spectra have to be different \cite{Weger}. Consequently one of the spectrum is emission spectrum. It is 
evident, taking into consideration the energy conservation law, that emission, corresponding to ASR phenomenon,  is observed by $\frac{dH_0}{dt} < 0$. It is understandable, that by the registration in CF-mode the process of accumulation of resonance phonons, for which the EM-field is the main heat reservoir, will be not realised. It is the consequence of effective back process, since  resonance phonons have the same energy with resonance photons and the only ESR responce can be registered in full accordance with experiment, see Figure 9.   

Although the emission, Figures 6, 8, is coherent, it is not maser effect, since the amplification is absent (maser effect seems to be in principle impossible for two level system). Then the hysteresis effect (compare Figure 5 and 7), which is observed by registration in AF-mode, can be explained in a natural way like to Stokes luminescence, where the emission spectrum is shifted to more low frequences in comparison with excitation spectrum to be consequence of interaction with nonresonance phonons. Given conclusion is confirmed by the study of the dependence of the Stokes shift value in ESR-ASR spectra of anthracite sample,  Figure 11, and  by the study of the dependence of
the effective linewidth in ESR spectra for both AF and CF registration modes, Figure 12,   on the amplitude of 100 kHz high frequency modulation of static magnetic field. Really passage velocity through the resonance is determined by product of $\omega_m H_m$, where $\omega_m$ is modulation frequency value. It means, that by passage velocity increase the direct  interaction of spin-photon subsystem with nonresonance phonons [which seems to be playing the same role, that the interaction with phonons by Stokes shift formation in luminescence studies] can  be "turned off", then emission and absorbtion spectra have to be registered without Stokes shift. Therefore 
the dependence of the Stokes shift value on the amplitude of 100 kHz high frequency modulation of static magnetic field has to to have steplike view and Stokes shift has to  vanish at high modulation amplitude side of given step. It really takes place, see Figure 11. Moreover Figure 11 allows to estimate the lifetime of coherent state in spin-photon subsystem from the $H_m$-value (in frequency units) in inflection of given curve. 
It gives the value $T_s = 1.3\times 10^{-6}$s, which gets to typical transverse $T_2$-value  range for many systems, described by classical Bloch equations. 
The possibility to "turn off" the direct  interaction of spin-photon subsystem with nonresonance phonons  by AF-mode registration means on the other hand, that the dependences of
the effective linewidth in ESR spectra by AF-mode registration and CF-mode registration have to be qualitatively different.  In AF-mode at small modulation amplitudes  the direct magnetic dipole interaction of spin subsystem in coherent state between the centers themselves takes place and it will lead to broadening of the registered ESR line, at that it seems to be evident that linewidth value has to increase with  the moving off  to the left and to decrease (without taking into account the modulation broadening) with  the moving off  to the right from  the $H_m$-value, corresponding  inflection in Figure 11 value (0.274 G).  It will be true, if genuine linewidth value is more than 0.274 G. It  is evident from CF-mode registration, that genuine linewidth value is less than 0.274 G, it is equal to 0.18 G. It means, that linewidth at inflection (that is at $H_m = 0.274 G$) will be modulationary broadened. The competition of  modulation broadening and magnetic dipole averaging processes (like to those ones in liquid systems) shifts the minimum position to the value $\approx 0.4 G$, see Figure 12, in AF-mode dependence of
the effective linewidth. Taking into account the value of  modulation broadening at  $\approx 0.4 G$, obtained from CF-mode dependence of
the effective linewidth, we can also evaluate the value of $T_s$. It is equal to  $T_s = 1.36 \times 10^{-6}$ s. Some difference of the value of $T_s$, obtained from experimental dependences, presented in Figures 11 and 12, is in limits of experimental inaccuracy, estimated to be equal to $10^{-7}$s. The value of $T_s$ allows to evaluate the range for the time of spin polarization transfer, corresponding to so called spin diffusion time $T_D$ by Dyson effect in metals. In the case of normal skin effect the value of $\frac{A}{B} = 23.4 \pm 0.5 $ means, that  $\frac{T_D}{T_s} < 10^{-4}$ and consequently $T_D < 1.3 \times 10^{-10}s$.

 The key to structure of ESR centers  can be found by the analysis of the central part of ESR spectra by  registration with automatic microwave frequency adjustment, presented in Figures 7 and 8. It is seen from Figure 7, that Rabi oscillation process is also takes place, although spectral resolution is not very good, since the modulation amplitude is coinciding with linewidth. The most interesting is the shape of background line. In both the spectra the shape of background line has kink form and can be be approximated by tanh-function. It can mean, that the model of the ESR centers has to be similar to Su-Schrieffer-Heeger (SSH) model of topological solitons in trans-polyacetylene \cite{Heeger_1988}. The g-factor value, which  is very near to g-factor value of neutral solitons  in trans-polyacetylene testifies in  favour of given proposal. It can mean, that structural element of anthracite includes carbon backbone of  trans-polyacetylene. At the same time anthracite is structurally disordered material.  It means, that the  solitons has to be pinned and they have to give usual ESR spectra. To explain the appearance of kinks, at that it is substantionally, that they are characterized by different parameters for absorption and emission spectra, and the possibility to register Rabi oscillations, we have to consider joint \{EM-field + magnetic dipole + resonance phonon field\} system. EM-field and resonance phonon field produce radiation communication in the direction of EM-field propagation and in time. It means that we have 1D chain, at that the chains  of two kinds are produced in time - with the sequence - absorption time segment - emission time segment and with reversal sequence. It is substantionally, that given time segments are different in their values. In fact the dimerisation take place like to trans-polyacetylene, however instead space z-axis along t-axis. It is in agreement with known equality of rights of coordinates in Minkowski space. In  other words the time has to be considered being to be quantized in given case. The junction of given time ordered chains produces time topological violations of two kinds. The first topological violation can effectively absorb photons and emit resonance phonons, the second topological violation has reversal properties. So, we come to the model of instanton, which is like to some extent to SSH-soliton model. It is understandable, that the  instantons of both kinds can free, that is without additional activation energy, move along t-axis like to SSH-solitons along space z-axis. The very short value of  $T_D < 1.3 \times 10^{-10} s$ is strong experimental support in the favour of the model proposed. Given model allows to explain the rather narrow ESR lines in structurally disordered materials like to sample studied, at that they are in more than 50 times narrower in comparison with ESR lines in  very good ordered related material - carbon nanotubes, produced by high energy implantation of diamond single crystals \cite{Ertchak}, \cite{Ertchak_Stelmakh}. The instantons seems to be two level noninteracting between themselves objects. Consequently they characterised by discrete energetic structure and the splitting into energetic band does not taking place, so we see that instanton model explain the appearance of narrow lines in anthracites in a natural way. It is the first experimental detection of instantons (to our knowledge).
The known instantons in the literature are considered to be a special kind of vacuum oscillations in gluon field theory and also in gravitation field theory \cite{Radjaraman}. They have quite another origin.
 
Therefore we have observed new kind of coherent states. They are modes of resonance phonon field. The coherence time value in the samples studied is evaluted to be equal  at room temperature $\sim 10 ms$, at that at relativelly low microwave power and it will  increase with power increase, achieving the maximal known values at present and even more. Substantial advantage for possible applications is that, that the responce on external EM-field is very strong. At the same time the detection of single spin is rather  technically difficult task. The H-field dependence allows to separate the phonon modes, the dependence on H-field direction seems to be also essential for practical applications.

\section{Conclusions} QED - QPFT model for  dynamics of spectroscopic transitions in 1D multiqubit  exchange coupled  system  is generalized by taking into account the earlier proof \cite{Yearchuck_Yerchak_Dovlatova}, that spin vector is  quaternion vector of the state of any quantum systen in Hilbert space defined  over quaternion ring and consequently all the spin components has to be taken into account. New quantum  phenomenon is predicted. The prediction results from the structure of  the equations derived and it consists in the following. The coherent system of the resonance phonons, that is,  the phonons with the energy, egualed to resonance photon  energy can be formed by resonance, that can lead to appearance along with   Rabi oscillations determined by spin (electron)-photon coupling with the frequency $\Omega^{RF}$ of Rabi oscillations determined by spin (electron)-phonon coupling with the frequency $\Omega^{RPh}$. In other words QED - QPFT model predicts the oscillation character of quantum relaxation, that is quite different character in comparison with phenomenological and semiclassical Bloch models.
  Moreover if  absolute value of  electron-phonon coupling constant $|\lambda^{\pm}_{\vec q l}|$, which characterisires  the interaction with    $S^+$- and $S^-_j$ components of the spin of jth chain unit is less, than electron(spin)-photon coupling constant $g$,  the model predicts that the second quantum Rabi oscillation process will be observed by stationary state of two subsystems \{EM-fied + magnetic (electric) dipoles\}, and it  will be registered in quadrature with the first quantum Rabi oscillation process. The second quantum Rabi oscillation process is governed by  the formation of the coherent system of the resonance phonons. Therefore along with absorption process of EM-field energy the coherent emission process takes place. Both the quantum Rabi oscillation processes can be time-shared. For the case of radiospectroscopy it corresponds to the possibility of the simultaneous observation  along with (para)magnetic spin resonance the acoustic spin resonance. The second (acoustic) quantum Rabi oscillation process can be detected even  by stationary spectroscopy methods. The model of new kind instantons is  proposed. All the predictions are experimentally confirmed for radiospectroscopy.  It has been found in particular, that the lifetime of coherent state of collective subsystem of resonance phonons in anthracites of medium-scale metamorphism  is very long and even by room temperature it is evaluated in $\sim 10$ ms. The phenomenon of  the formation of the coherent system of the resonance phonons can find the  number of practical applications, in particular it can be used by elaboration of logic quantum systems including quantum computers and quantum communication systems.

\end{document}